\documentclass[conference]{IEEEtran}
\IEEEoverridecommandlockouts
\usepackage{cite}
\usepackage{amsmath,amssymb,amsfonts}
\usepackage{algorithmic}
\usepackage{graphicx}
\usepackage{textcomp}
\usepackage{xcolor}
\usepackage{algorithm}
\usepackage{algorithmic}
\usepackage{fontspec}
\usepackage{subcaption}

\AtBeginDocument{
  \setlength{\abovedisplayskip}{4pt}
  \setlength{\belowdisplayskip}{4pt}
  \setlength{\abovedisplayshortskip}{2pt}
  \setlength{\belowdisplayshortskip}{2pt}
  \setlength{\jot}{2pt}
}

\usepackage{fontspec}
\makeatletter
\renewcommand{\ALG@name}{Algorithm}
\def\BibTeX{{\rm B\kern-.05em{\sc i\kern-.025em b}\kern-.08em
    T\kern-.1667em\lower.7ex\hbox{E}\kern-.125emX}}
    
\begin{document}

\title{Sum Rate optimization for RIS-Aided RSMA system with Movable Antenna
}

% 1\textsuperscript{st} 2\textsuperscript{nd} 3\textsuperscript{rd} 
\author{
	\IEEEauthorblockN{
		Mingyu Hu\IEEEauthorrefmark{*},
		Nan Liu\IEEEauthorrefmark{+} and
		Wei Kang\IEEEauthorrefmark{*}
	\IEEEauthorblockA{\IEEEauthorrefmark{*}School of Information Science and Engineering, Southeast University, Nanjing, China 210096}
	\IEEEauthorblockA{\IEEEauthorrefmark{+}National Mobile Communications Research Laboratory, Southeast University, Nanjing, China 210096}
	\IEEEauthorblockA{\{myuhu, nanliu, wkang\}@seu.edu.cn}
}
%\author{\IEEEauthorblockN{Chen Xu, Wei Kang}
%\IEEEauthorblockA{\textit{School of Information Science and Engineering} \\
%\textit{Southeast University}\\
%Nanjing, China \\
%cxuze@seu.edu.cn, wkang@seu.edu.cn}
%\and
%\IEEEauthorblockN{Nan Liu}
%\IEEEauthorblockA{\textit{National Mobile Communications Research Laboratory} \\
%\textit{Southeast University}\\
%Nanjing, China \\
%nanliu@seu.edu.cn}
% \and
% \IEEEauthorblockN{3\textsuperscript{rd} Wei Kang}
% \IEEEauthorblockA{\textit{School of Information Science and Engineering} \\
% \textit{Southeast University}\\
% Nanjing, China \\
% wkang@seu.edu.cn}
% \and
% \and
% \and
\thanks{This work is partially supported by the National Natural Science Foundation of China under Grants $62361146853$, and $62371129$, and~the Research Fund of the National Mobile Communications Research Laboratory, Southeast University, under Grant 2026A05.}
}

\DeclareRobustCommand*{\IEEEauthorrefmark}[1]{
	\raisebox{0pt}[0pt][0pt]{\textsuperscript{\footnotesize\ensuremath{#1}}}}

\maketitle

\begin{abstract}
% Rate-Splitting Multiple Access (RSMA) is taken as a key enabling technique for sixth-generation (6G) wireless systems for its powerful interference management. Reconfigurable Intelligent Surface (RIS) can effectively shape the wireless propagation to match the environment and improve communication performance. However, in conventional RSMA–RIS architectures, the antenna elements are fixed, which underutilizes spatial degrees of freedom and hence constrains system performance. To address this limitation, we propose a movable-antenna (MA) assisted RSMA–RIS framework and formulate a sum-rate maximization problem that jointly optimizes the transmit beamforming matrix, the RIS reflection matrix, the common-rate partition, and the MA positions. The original problem is transformed by employing the fractional programming (FP) method, and a closed-form solution for the common rate splitting is derived. Leveraging the Karush–Kuhn–Tucker (KKT) conditions, we obtain iterative updates for the Lagrange multipliers together with a closed-form expression for the beamforming matrix. We then develop an update rule for the RIS reflection matrix via the dual problem, and finally determine the optimal antenna locations using a gradient-ascent procedure. Numerical results indicate that, in the presence of RIS assistance, incorporating MAs yields additional performance improvements for both space-division multiple access (SDMA) and RSMA, with gains of approximately 33.3\% for SDMA and 35.6\% for RSMA, respectively.

Rate-Splitting Multiple Access (RSMA) is a key enabling technique for sixth-generation (6G) wireless systems due to its powerful interference management, and Reconfigurable Intelligent Surface (RIS) improves communication performance by shaping wireless propagation. However, conventional RSMA--RIS architectures employ fixed antennas, limiting spatial degrees of freedom and system performance. To address this, we propose a movable-antenna (MA) assisted RSMA--RIS framework and formulate a sum-rate maximization problem that jointly optimizes the transmit beamforming matrix, RIS reflection matrix, common-rate partition, and MA positions. After yielding a closed-form solution for common rate splitting, the problem is transformed via fractional programming (FP). Using Karush--Kuhn--Tucker (KKT) conditions, we give iterative updates for Lagrange multipliers and beamforming matrix, obtain the RIS reflection matrix via the dual problem, and determine optimal antenna positions via gradient ascent. Numerical results show that with the existence of RIS, integrating MA yields additional gains of approximately 33.3\% for SDMA and 35.6\% for RSMA.

\end{abstract}


\section{Introduction}

Movable antenna (MA) has recently emerged as a promising technique delivering substantial performance gains in wireless systems~\cite{zhu2023modeling}. By connecting each radiating element to the RF chain via a flexible coaxial cable~\cite{nithya2022design}, MA enables adaptive repositioning within a multi-wavelength region, thereby expanding spatial degrees of freedom, supporting high-quality beamforming, and reducing hardware cost relative to antenna selection (AS) schemes. Unlike conventional fixed-position antennas (FPAs) restricted to 1D/2D layouts, MA exploits full 3D spatial flexibility. These advantages have been validated across diverse scenarios. Demonstrated in \cite{ma2023mimo}, the deployment of movable antenna (MA) yields significant capacity gains in multiple-input multiple-output (MIMO) systems. The authors of \cite{yan2025movable} extend the performance enhancement of MA for MIMO systems to the case of statistical channel conditions. Moreover, the performance enhancement of MAs in multiple access channels (MACs) was analyzed in \cite{zhu2023movable}. Additionally, MAs have also been empirically verified to improve system performance in various other communication scenarios, such as enhancing the efficiency of mobile edge computing (MEC) systems \cite{cang2025movable} and enabling satisfactory secrecy performance in covert communication systems \cite{liu2024movable}. %Furthermore, the authors of \cite{wu2025movable} investigated MA-assisted reconfigurable intelligent surface–integrated sensing and communication (RIS–ISAC) systems, showing that MAs can reshape the channel characteristics and substantially improve the overall channel gain of ISAC systems.

% Rate-Splitting Multiple Access (RSMA) is a multiple-access paradigm poised for next-generation wireless networks. It partitions each user’s message into a common part and a private part, which are encoded into a common stream and private streams, respectively\cite{mao2018rate}. Users first decode the common stream by treating all private streams as noise. After canceling the common stream, which partially mitigates multiuser interference, each user decodes its own private message while treating the remaining private streams as noise.
% Various RSMA architectures have been developed and studied in which 1-layer RSMA has been demonstrated to deliver strong performance and is widely adopted\cite{hao2015rate}. Also, RSMA demonstrates robust interference management capabilities under various channel conditions. Specifically, in the limiting cases of weak or strong inter-user interference, it automatically adjusts the split between common and private information and degenerates into space-division multiple access (SDMA) or non-orthogonal multiple access (NOMA), enabling flexible operation across scenarios\cite{mao2022rate}. Moreover, RSMA affords higher degrees of freedom than SDMA and NOMA, leading to faster growth of user rates with the signal-to-noise ratio (SNR)\cite{hao2017achievable}\cite{piovano2017optimal}.
	
Rate-Splitting Multiple Access (RSMA) is a promising multiple-access scheme for next-generation wireless networks. It splits each user’s message into a common part and a private part, encoded into one common stream and multiple private streams, respectively~\cite{mao2018rate}. Each user first decodes the common stream by treating all private streams as noise, and then decodes its own private message after removing the common stream. Among various RSMA architectures, 1-layer RSMA is widely adopted due to its strong performance~\cite{hao2015rate}. RSMA also enables robust interference management under diverse channel conditions. In particular, under weak or strong inter-user interference, it can adapt the common-private split and reduce to SDMA or NOMA, respectively~\cite{mao2022rate}. Moreover, RSMA achieves higher degrees of freedom than SDMA and NOMA, yielding faster user-rate growth with the signal-to-noise ratio (SNR)~\cite{hao2017achievable,piovano2017optimal}.

	Reconfigurable intelligent surface (RIS) is a kind of metasurface technology consisting of a two-dimensional array of reflective elements, each of which can independently adjust the phase of the incident signal according to the instantaneous channel, thereby synthesizing highly directive reflected beams and enhancing system performance—especially when the line-of-sight (LoS) path between the base station (BS) and users is blocked or severely attenuated by obstacles. Extensive investigations on RIS have been performed across diverse deployment scenarios like in \cite{wu2019intelligent,huang2019reconfigurable,yang2021energy}.
	
	Typically, RSMA outperforms SDMA and NOMA when practical channels between the base station and users fall between the extremes of orthogonality and full alignment\cite{mao2018rate}. Since both MA and RIS can actively shape the effective channel, their benefits are expected to be more pronounced for RSMA. To the best of our knowledge, the performance gains of MAs for RSMA–RIS systems remain unexplored. Motivated by this gap, we propose a downlink multiuser MISO scheme with MA-assisted RSMA–RIS, wherein we maximize the sum rate by jointly optimizing the transmit beamforming matrix, the RIS reflection matrix, the common-rate partition, and the MA positions. The problem is recast via fractional programming (FP) into a more tractable form. Efficient updates are then derived using the Karush–Kuhn–Tucker (KKT) conditions and associated dual problem. Finally the antenna positions are obtained through a gradient-ascent procedure. Our main contributions are summarized as follows:
	
	\begin{itemize}
		\item To the best of our knowledge, this is the first work to investigate the performance gains of MA for RSMA--RIS system. We formulate a sum-rate maximization that jointly optimizes the transmit beamforming matrix, the RIS reflection matrix, the common-rate allocation, and the MA positions.
		
		\item We propose an effective alternating optimization (AO) iterative algorithm to solve the formulated optimization problem. Specifically, by decomposing the initial problem into several subproblems, each variables can be updated via corresponding approaches.
		
		% the variables are we derive a closed-form solution for the common-rate allocation and, via FP method, reformulate the original problem into a more tractable equivalent form. Leveraging the KKT condition, we obtain iterative updates for the beamforming matrix. Moreover, we develop an update rule for the RIS reflection matrix through associated dual problem. Finally, gradient-ascent scheme is employed to yield near-optimal MA positions. 
		
		\item Numerical results show that, MA can improve the performance of RSMA-RIS system compared with FPA. Also, the algorithm proposed in this work can effectively solve the initial problem and outperforms the benchmark algorithms.     %Moreover, the MA-induced improvement for RSMA exceeds that for SDMA, and this advantage becomes more pronounced as the number of users increases.
	\end{itemize}

	% Neural 方法	
\section{System Model}
In this paper, we consider the system shown in Fig.~1, where the BS is equipped with $M$ movable antennas and adopts 1-layer RSMA to split user messages. The BS simultaneously serves $K$ far-field single-antenna users in the presence of obstacles, which create non-line-of-sight (NLOS) paths. For simplicity, each antenna is allowed to move horizontally via flexible cables. In addition, a RIS with $N$ reflecting elements is deployed to enhance communication performance.

	\subsection{Channel Model}
	In this work, the field response model in \cite{ma2023mimo} is adopted for channel modeling. Let $\mathcal{M}=\{1,\cdots,M\}$ denote the set of MAs, with positions $\mathbf{x}=[x_1,x_2,\dots,x_M]$, where $x_m$ is the position of the $m$-th MA. The end-to-end channel consists of three components. First, we assume that the channel state information (CSI) of the RIS--user link, denoted by $\mathbf{h}_{r,k}\in\mathbb{C}^{N\times 1}$, is available at the BS. We consider a geometric channel model with $L_t$ transmit paths from the BS and $L_r$ receive paths at the RIS. The far-field response vector (FRV) from the BS to the RIS is defined as $\mathbf{a}(x_i)=[e^{j\frac{2\pi}{\lambda}x_i\cos\theta_1^t},\cdots,e^{j\frac{2\pi}{\lambda}x_i\cos\theta_{L_t}^t}]^T\in\mathbb{C}^{L_t\times 1}$, where $\theta_i^t$, $i\in\{1,2,\ldots,L_t\}$, denotes the transmit angle of the $i$-th BS propagation path. Stacking all FRVs yields the field response matrix (FRM) $\mathbf{A}(\mathbf{x})=[\mathbf{a}(x_1),\mathbf{a}(x_2),\cdots,\mathbf{a}(x_M)]\in\mathbb{C}^{L_t\times M}$. Similarly, the receive steering vector of the RIS is defined as $\mathbf{b}_i=[b_{i,1},\cdots,b_{i,L_r}]^T$, where $b_{i,j}=e^{j\frac{2\pi}{\lambda}(x_{r,i}\sin\theta_j^r\cos\phi_j^r+y_{r,i}\cos\theta_j^r)}$, $(x_{r,i},y_{r,i})$ is the coordinate of the $i$-th RIS element, and $\theta_i^r$ and $\phi_i^r$ are the corresponding elevation and azimuth angles at the RIS receive side. By stacking these vectors, we obtain $\mathbf{B}=[\mathbf{b}_1,\mathbf{b}_2,\cdots,\mathbf{b}_N]\in\mathbb{C}^{L_r\times N}$. The equivalent BS--RIS channel is $\mathbf{H}_{b,r}(\mathbf{x})=\mathbf{B}^H\boldsymbol{\Sigma}\mathbf{A}(\mathbf{x})$, where $\boldsymbol{\Sigma}\in\mathbb{C}^{L_r\times L_t}$ is the BS--RIS fading matrix, and each entry $\boldsymbol{\Sigma}[i,j]$ denotes the fading coefficient between the $i$-th receive path and the $j$-th transmit path, following a complex Gaussian distribution. Similarly, the BS--user response matrix is $\mathbf{A}_k(\mathbf{x})=[\mathbf{a}_k(x_1),\cdots,\mathbf{a}_k(x_M)]$, where $\mathbf{a}_k(x_i)=[e^{j\frac{2\pi}{\lambda}x_i\cos\theta_{k,1}^t},\cdots,e^{j\frac{2\pi}{\lambda}x_i\cos\theta_{k,L_t}^t}]^T$ and $\theta_{k,i}^t$ denotes the transmit angle of the $i$-th BS--user $k$ path. Accordingly, the BS--user channel is $\mathbf{h}_{b,k}^H=\mathbf{1}^H\boldsymbol{\Sigma}_k\mathbf{A}_k(\mathbf{x})$, where $\mathbf{1}$ is an all-one row vector and $\boldsymbol{\Sigma}_k\in\mathbb{C}^{L_r\times L_t}$ is the BS--user $k$ fading matrix. Each entry $\boldsymbol{\Sigma}_k[i,j]$ denotes the fading coefficient between the $i$-th receive path and the $j$-th transmit path and also follows a complex Gaussian distribution. Therefore, the combined equivalent channel from the BS to user $k$ is $\mathbf{h}_k^H(\mathbf{x})=\mathbf{h}_{r,k}^H\boldsymbol{\Phi}\mathbf{H}_{b,r}(\mathbf{x})+\mathbf{h}_{b,k}^H$, where $\boldsymbol{\Phi}=\operatorname{diag}([\phi_1,\cdots,\phi_N])$, and $\phi_i$ is the reflection coefficient of the $i$-th RIS element with $|\phi_i|=1$ for all $i$, where $|\cdot|$ denotes the magnitude.

	\subsection{Signal Model}
	In a 1-layer RSMA, the message for user $k$ which is denoted as $W_k$ is divided into a common part $W_{c,k}$ and a private part $W_{p,k}$. %The private part of each user is encoded into $\{s_1,s_2,...,s_K \}$ 
	The private part of each user is encoded into an independent private stream $s_k$ for $k = 1,\cdots,K$. The common parts of each user are combined into one part and consequently encoded into $s_c$. It is assumed that each stream in $\mathbf{s}=\{s_1,s_2,...,s_K,s_c \}$ satisfies $\mathbb{E}[s_i \bar{s}_i]=1$ and $\mathbb{E}[s_i \bar{s}_j]=0, \forall i \neq j$\cite{fang2024rate}. $\bar{(\cdot)}$ denotes the conjugate. We denote the beamformer of the $k$-th private stream and the common stream as $\mathbf{w}_k$ and $\mathbf{w}_c$, respectively. Then we can define the beamforming matrix $\mathbf{W}$ as $\mathbf{W}=[\mathbf{w}_1,\cdots,\mathbf{w}_K,\mathbf{w}_c] \in \mathbb{C}^{M \times (K+1)}$. Therefore, the received signal at user $k$ can be expressed as:
	\begin{align}\label{receivesignal}
		y_k=\mathbf{h}_k^H(\mathbf{x}) \mathbf{W} \mathbf{s}+n_k,
	\end{align} 
	 where $n_k \sim \mathcal{CN}(0,\sigma_{k}^2)$ is the receiver noise of the user $k$. According to the 1-layer RSMA scheme, the user decodes the common stream while regarding other streams as interference. After canceling the common part from the stream, users proceed to decode their private parts. The signal-to-interference-plus-noise ratio (SINR) for decoding the common part at user $k$ can be expressed as:
	\begin{align}\label{SINRCK}
		\operatorname{SINR}_{c,k}=\frac{|\mathbf{h}_k^H(\mathbf{x})\mathbf{w}_c|^2}{\sum_{i=1}^K |\mathbf{h}_k^H(\mathbf{x})\mathbf{w}_{{i}}|^2+\sigma_k^2}
	\end{align}  
	Consequently, we can get the maximum rate of user $k$ for decoding common stream as:
	\begin{equation}\label{RCK}
	R_{c,k}=\operatorname{log}_2(1+\operatorname{SINR}_{c,k})
	\end{equation}
	To ensure that each user can correctly decode the common stream $s_c$, the sum of the allocated common rates, $\sum_{k} r_{c,k}$, should not exceed $\min_k\{R_{c,k}\}$, where $r_{c,k}$ is the common rate allocated to the user $k$. Then the SINR for decoding the private stream at the user $k$ is given by:
	
	\begin{align}\label{SINRK}
		\operatorname{SINR}_k=\frac{|\mathbf{h}_k^H(\mathbf{x})\mathbf{w}_k|^2}{\sum_{i=1,i\neq k}^K |\mathbf{h}_k^H(\mathbf{x})\mathbf{w}_i|^2+\sigma_k^2}
	\end{align}
	Then, the maximum rate for decoding the private stream at user $k$ is denoted as:
	\begin{equation}
	R_k = \operatorname{log}_2(1+\operatorname{SINR}_k)
	\end{equation}
	% \begin{align}\label{RK}
	% 	R_k = \operatorname{log}_2(1+\operatorname{SINR}_k) 
	% \end{align}

	\begin{figure}[t]
		\centering
		\vspace{1mm} % 可微调上下间距
		\includegraphics[width=0.9\columnwidth, keepaspectratio, trim=0 0 0 0, clip]{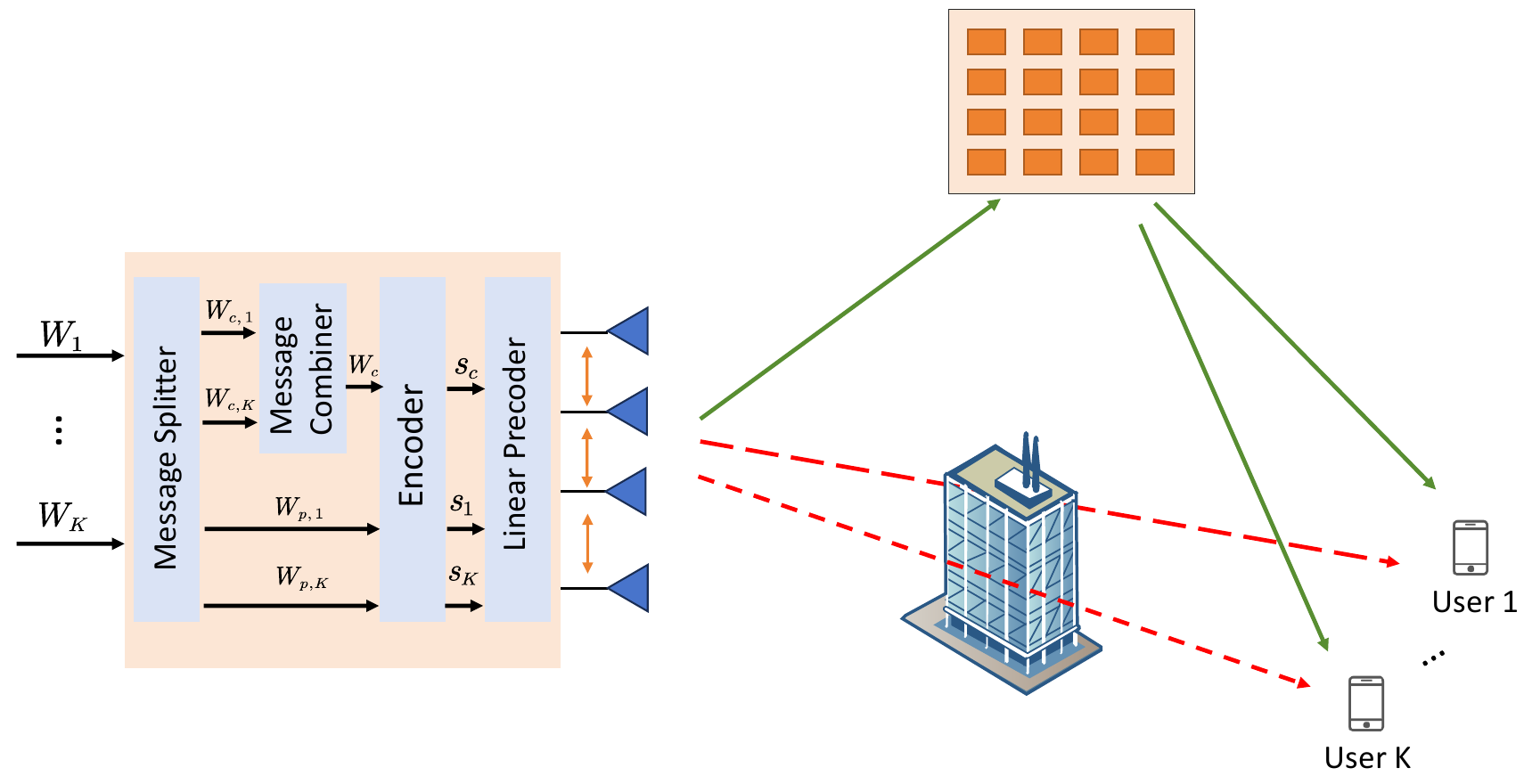}
		\caption{System model of an MA-aided RSMA-RIS system.}
		\label{fig:system_model}
		\vspace{-2mm} % 调整与正文的距离，可根据实际效果修改
	\end{figure}

	\subsection{Problem Formulation}
	To investigate whether MA can enhance system performance in the RSMA-RIS system, we formulate an optimization problem that aims to maximize the sum of both the private and common achievable rates of all users. The optimization problem is formulated as:
	 \begin{subequations}
	 	\begin{align}
	 	(\mathbf{P1}) 	\max_{\mathbf{x}, \mathbf{W}, \boldsymbol{\Phi}, \{ r_{c,k} \}} 
	 		& \quad R=\sum_{k=1}^{K} (R_{k} + r_{c,k}) \label{eq:opt_problem_obj} \\
	 		\text{s.t.} \quad 
	 		& \mathrm{tr}\left(\mathbf{W}^H \mathbf{W}\right) \leq P_T, \label{eq:opt_problem_constraint1} \\
			&R_k \geq \iota,\quad \forall k \label{P1rate}\\
	 		& \sum_{k=1}^{K} r_{c,k} \leq R_{c,k}, \quad \forall k, \label{eq:opt_problem_constraint2} \\
	 		& r_{c,k} \geq 0, \quad \forall k, \label{eq:opt_problem_constraint3} \\
	 		& |\phi_i|^2 = 1, \quad \forall i, \label{eq:opt_problem_constraint4} \\
	 		& X_{min}\leq x_i \leq X_{max}, \label{region}\\
	 		& \|x_i - x_j\|_2 \geq D_0, \quad \forall i \neq j. \label{eq:opt_problem_constraint5}
	 	\end{align}
	 \end{subequations}

	 Constraint \eqref{eq:opt_problem_constraint1} is the power constraint with $P_T$ denoting the power budget of the BS, \eqref{P1rate} is the private rate constraint to ensure the fairness with $\iota$ the threshold for private rate, \eqref{eq:opt_problem_constraint2} is the common rate constraint to make sure that every user can successfully encode the common stream, \eqref{eq:opt_problem_constraint3} represents the non-negativity constraint of the common rate, \eqref{eq:opt_problem_constraint4} represents the constant-modulus constraint imposed on each RIS element, \eqref{region} represents the minimum inter-antenna spacing constraint with $X_{min}$ and $X_{max}$ the limits of region for MA, \eqref{eq:opt_problem_constraint5} is the minimum distance between antennas with $D_0$ the minimum inter-antenna spacing between any two antennas.
	 
	 Due to the non-convex nature of \eqref{eq:opt_problem_obj}, \eqref{P1rate}, \eqref{eq:opt_problem_constraint4} and \eqref{eq:opt_problem_constraint5}, as well as the strong coupling among the optimization variables, problem $(\mathbf{P1})$ is a highly non-convex optimization problem. In the subsequent sections, we develop an efficient alternating optimization (AO) algorithm to solve this problem.

\section{PROPOSED SOLUTION}
	In this section, we propose an efficient AO algorithm to solve problem $(\mathbf{P1})$. Specifically, by fixing a subset of variables and optimizing the remaining ones, the original problem is decomposed into several tractable subproblems. Moreover, closed-form solutions are derived in some subproblems, which reduces the overall computational complexity.

	\subsection{Updating Allocated Common Rates}
	When all optimization variables except for the allocated common rate are fixed, the original problem can be transformed into:
	% \begin{subequations}\label{eq:P2}
	% 	\begin{align}
	% 		(\mathbf{P2})\quad \max_{\{ r_{c,k} \}} \quad 
	% 		& \sum_{k=1}^{K} r_{c,k} \label{eq:P2_obj} \\[1ex]
	% 		\text{s.t.} \quad 
	% 		& \sum_{k=1}^{K} r_{c,k} \leq R_{c,k}, \quad \forall k, \label{eq:P2_constraint1} \\[1ex]
	% 		& r_{c,k} \geq 0, \quad \forall k, \label{eq:P2_constraint2}
	% 	\end{align}
	% \end{subequations}
	\begin{equation}
		(\mathbf{P2})~ \max_{\{ r_{c,k} \}}~\sum_{k=1}^{K} r_{c,k} \quad \text{s.t.}~\eqref{eq:opt_problem_constraint2},\eqref{eq:opt_problem_constraint3}\nonumber
	\end{equation}
	where constraint \eqref{eq:opt_problem_constraint2} can be equivalently rewritten as $\sum_{k=1}^K r_{c,k} \leq \operatorname{min}_k\{R_{c,k}\}$. This reformulation indicates that the sum of the common-rate allocations assigned to all users cannot exceed the minimum achievable common rate among all users. Since the objective function is monotonically increasing with respect to $r_{c,k}$, its maximum value is attained when this constraint is satisfied with equality. In other words, at the optimum, the entire available common rate should be fully utilized. Therefore, the optimal common-rate allocation can be chosen as an equal distribution among all users, i.e., $r_{c,k}^*=\frac{1}{K}\operatorname{min}_k\{R_{c,k}\},\forall k$, where $(\cdot)^*$ denotes the optimal value.

	\subsection{Transformation Based on FP}
	In this section, the FP method\cite{shen2018fractional} is employed to transform the intractable non-convex objective function \eqref{eq:opt_problem_obj} and constraints \eqref{eq:opt_problem_constraint2} into a more tractable form. By substituting the optimal common rate into $(\mathbf{P1})$ and applying the FP-based reformulation, an equivalent problem, denoted as $(\mathbf{P3})$, can be obtained as: 
	\begin{subequations}\label{eq:P2_FP}
		\begin{align}
		\mathbf{(P3)}	\max_{\mathbf{W}, \boldsymbol{\Phi}, \mathbf{x}, \boldsymbol{\mu}, \boldsymbol{\epsilon}, \boldsymbol{\gamma}, \mathbf{v}, y} \quad
			& \sum_{k=1}^{K} \Psi_k (\boldsymbol{\mu}, \boldsymbol{\epsilon}, \mathbf{W}, \mathbf{x}, \boldsymbol{\Phi}) + y \label{eq:P2FP_obj} \\
			\text{s.t.} \quad 
			& T_k (\boldsymbol{\gamma}, \mathbf{v}, \mathbf{W}, \boldsymbol{\Phi}, \mathbf{x}) \geq y, \quad \forall k, \label{P3commonrate}\\
			&\Psi_k(\boldsymbol{\mu}, \boldsymbol{\epsilon}, \mathbf{W}, \mathbf{x}, \boldsymbol{\Phi})\geq \iota,\quad \forall k,\label{P3rate}\\
			&\eqref{eq:opt_problem_constraint1},\eqref{region},\eqref{eq:opt_problem_constraint5} \nonumber
		\end{align}
	\end{subequations}
where $y$ is a slack variable introduced to replace $\min_{k}\{R_{c,k}\}$\cite{fang2024rate}, and $\boldsymbol{\mu}=[\mu_1,\cdots,\mu_K]^T$, $\boldsymbol{\epsilon}=[\epsilon_1,\cdots,\epsilon_K]^T$, $\boldsymbol{\gamma}=[\gamma_1,\cdots,\gamma_K]^T$, $\mathbf{v}=[v_1,\cdots,v_K]^T$ are auxiliary variables introduced by the FP technique. The functions $\Psi_k$ and $T_k$ are given in \eqref{psik} and \eqref{tk}.

	\subsubsection{Updating Auxiliary Variables $\boldsymbol{\mu}$, $\boldsymbol{\gamma}$, $\boldsymbol{\epsilon}$, $\boldsymbol{v}$}
	
	It is noteworthy that when other variables are fixed, equations \eqref{psik} and \eqref{tk} are concave functions with respect to the auxiliary variables. Consequently, their updates can be obtained by taking the corresponding derivatives. %However, it should be emphasized that the derivations of $\boldsymbol{\mu}$ and $\boldsymbol{\gamma}$ are computed with respect to the Lagrangian-transformed functions prior to the FP reformulation\cite{shen2018fractional}, whereas the remaining two auxiliary variables are directly differentiated from equations \eqref{psik} and \eqref{tk}.
	The updates of auxiliary variables are given in (15)$\sim$(18) in \cite{11161312}. They are ignored here for simplicity.
	% \vspace{-0.6em}
	% \begingroup
	% \setlength{\abovedisplayskip}{3pt}
	% \setlength{\belowdisplayskip}{3pt}
	% \setlength{\abovedisplayshortskip}{2pt}
	% \setlength{\belowdisplayshortskip}{2pt}
	% \setlength{\jot}{1pt}
	% \begin{subequations}
	% \begin{align}
	% \label{updatemu*}
	% \mu_k^{*} &=
	% \frac{|\mathbf{h}_k^{H}(\mathbf{x}) \mathbf{w}_k|^2}
	% {\sum_{i \ne k}^{K} |\mathbf{h}_k^{H}(\mathbf{x}) \mathbf{w}_i|^2 + \sigma_k^2}\\
	% \label{updateepsilon*}
	% \epsilon_k^{*} &=
	% \frac{\sqrt{1 + \mu_k}\,\mathbf{h}_k^{H}(\mathbf{x}) \mathbf{w}_k}
	% {\sum_{i=1}^{K} |\mathbf{h}_k^{H}(\mathbf{x}) \mathbf{w}_i|^2 + \sigma_k^2}\\
	% \label{updategamma*}
	% \gamma_k^{*} &=
	% \frac{|\mathbf{h}_k^{H}(\mathbf{x}) \mathbf{w}_c|^2}
	% {\sum_{i=1}^{K} |\mathbf{h}_k^{H}(\mathbf{x}) \mathbf{w}_i|^2 + \sigma_k^2}\\
	% \label{updatev*}
	% v_k^{*} &=
	% \frac{\sqrt{1 + \gamma_k}\,\mathbf{h}_k^{H}(\mathbf{x}) \mathbf{w}_c}
	% {|\mathbf{h}_k^{H}(\mathbf{x}) \mathbf{w}_c|^2 + \sum_{i=1}^{K} |\mathbf{h}_k^{H}(\mathbf{x}) \mathbf{w}_i|^2 + \sigma_k^2}
	% \end{align}
	% \end{subequations}
	% \endgroup
	% \vspace{-0.8em}
	
	\subsubsection{Updating Beamforming Matrix $\mathbf{W}$ and Slack Variable $y$}
\vspace{-0.2em}
With other variables fixed, the problem $(\mathbf{P3})$ can be reformulated as:
\vspace{-0.3em}

{\setlength{\abovedisplayskip}{2pt}
\setlength{\belowdisplayskip}{2pt}
\setlength{\abovedisplayshortskip}{1pt}
\setlength{\belowdisplayshortskip}{1pt}
\setlength{\jot}{0pt}
\begin{subequations}
\begin{align}
(\mathbf{P3.1})~ \max_{\mathbf{W},y}\quad
& \sum_{k=1}^{K} \Psi_k(\mathbf{W}) + y \quad\text{s.t.}~\eqref{eq:opt_problem_constraint1},\eqref{P3commonrate},\eqref{P3rate} \nonumber
\end{align}
\end{subequations}
}

% \label{P3.1func}\\
% \text{s.t.}\quad
% & T_k(\mathbf{W}) \ge y, \quad \forall k, \label{P3.1commonrate}\\
% & \Psi_k(\mathbf{W}) \ge \iota, \quad \forall k, \label{P31rate}\\
% & \eqref{eq:opt_problem_constraint1} \nonumber
% \end{align}
% \end{subequations}
% \vspace{-0.8em}

This is a convex problem. Although tools such as CVX can solve it effectively, they incur higher complexity. Therefore, in this subsection, we adopt the algorithm in~\cite{fang2024rate} to solve the beamforming matrix via the KKT conditions. The update formula for the beamforming matrix $\mathbf{W}$ is given by:
{\setlength{\abovedisplayskip}{4pt}
\setlength{\belowdisplayskip}{4pt}
\setlength{\abovedisplayshortskip}{2pt}
\setlength{\belowdisplayshortskip}{2pt}
\setlength{\jot}{1pt}
\begin{subequations}\label{eq:KKT}
\begin{align}
	&\mathbf{w}_{c}^{*}
	= 
	\left(
	\sum_{k=1}^{K} \lambda_k^{*} |v_k|^2 \mathbf{h}_k(\mathbf{x}) \mathbf{h}_k^{H}(\mathbf{x}) + \kappa^{*} \mathbf{I}
	\right)^{-1} \nonumber\\
	&\quad \quad \quad \quad \quad \quad \quad \quad \quad\sum_{k=1}^{K} (\lambda_k^{*} \sqrt{1+\gamma_k}\, v_k \mathbf{h}_k(\mathbf{x}))
	 \label{eq:KKT_a} \\  % 原为2ex，略缩
	&\mathbf{w}_{m}^{*}
	=
	\left(
	\sum_{k=1}^{K} (|\epsilon_k|^2(1+\eta_k) + \lambda_k^{*} |v_k|^2) \mathbf{h}_k(\mathbf{x}) \mathbf{h}_k^{H}(\mathbf{x}) + \kappa^{*} \mathbf{I}
	\right)^{-1} \nonumber\\
	& \quad\quad\quad\quad\quad\quad\quad\quad2(1+\eta_m)\sqrt{1+\mu_m}\, \epsilon_m\, \mathbf{h}_m(\mathbf{x}), ~\forall m \neq c, 
	\label{eq:KKT_b}
	% &\quad\quad\quad\quad\sum_{k=1}^{K} \lambda_k^{*} = 1, \label{eq:KKT_c}\\[0.8ex]
	% &\quad\quad\quad\quad\lambda_k^{*}(y^{*} - T_k(\mathbf{W}^{*})) = 0, \quad \forall k, \label{eq:KKT_d}\\[0.8ex]
	% &\quad\quad\quad\quad\kappa^{*}(\mathrm{tr}(\mathbf{W^*}^{H}\mathbf{W^*}) - P_T) = 0, \label{eq:KKT_e}\\[0.8ex]
	% &\quad\quad\quad\quad\eta_k^*(\iota-\Psi_k(\mathbf{W}^*))=0, \quad \forall k,\label{KKT_rate}\\[0.8ex]
	% &\quad\quad\quad\quad\lambda_k^{*} \ge 0, \quad \kappa^{*} \ge 0,\quad \eta_k\geq 0, \quad \forall k \label{eq:KKT_f}
\end{align}
\end{subequations}
}where $\boldsymbol{\lambda}=[\lambda_1,\cdots,\lambda_K]^T$, $\kappa$ and $\boldsymbol{\eta}=[\eta_1,\cdots,\eta_K]$ are the Lagrange multipliers associated with constraints \eqref{P3commonrate} , \eqref{eq:opt_problem_constraint1} and \eqref{P3rate}, respectively. Among them, $\mathbf{I}$ is the identity matrix, equations \eqref{eq:KKT_a} and \eqref{eq:KKT_b} arise from stationary condition. 
	% while equations \eqref{eq:KKT_d} and \eqref{eq:KKT_e} correspond to the complementary slackness conditions. 
	According to the structure of problem $(\mathbf{P3.1})$, the optimal value of $y$ is $y^{*} = T_{z}(\mathbf{W})$,
	% \begin{equation}\label{optimaly}
	% 	y^{*} = T_{z}(\mathbf{W}).
	% \end{equation}
	where $z$ denotes the user index corresponding to $\min_k$\{$T_k$\}. Applying the fixed point iteration method in \cite{fang2024rate}, the update rules for the Lagrange multipliers can be derived as follows:
	\begin{subequations}
		\begin{align}
			\lambda_k^{[t+1]} 
			&= 
			\frac{T_z(\mathbf{W}) + \rho}
			{T_k(\mathbf{W}) + \rho} 
			\, \lambda_k^{[t]}, 
			\quad \forall k \ne z, \label{eq:lambda_update_a}\\
			\lambda_z^{[t+1]} 
			&= 
			\lambda_z^{[t]} 
			+ \sum_{k=1}^{K} 
			\left(
			\lambda_k^{[t]} 
			- 
			\frac{T_z(\mathbf{W}) + \rho}
			{T_k(\mathbf{W}) + \rho} 
			\, \lambda_k^{[t]}
			\right), \label{eq:lambda_update_b}\\
			\eta_k^{[t+1]}&=\frac{\Psi_k(\mathbf{W})+\rho}{\iota+\rho},\label{updateiota}\\
			\kappa^{[t+1]} 
			&= 
			\frac{\mathrm{tr}(\mathbf{W}\mathbf{W}^{H}) + \rho}
			{P_T + \rho}, \label{eq:lambda_update_c}
		\end{align}
	\end{subequations}
	where constant $\rho \geq 0$ is employed to enhance convergence stability by effectively reducing the step size. $(\cdot)^{[t]}$ and $(\cdot)^{[t+1]}$ are the value of $t$-th and ($t{+}1$)-th iteration respectively. Equation \eqref{eq:lambda_update_b} is designed to ensure that the KKT condition \eqref{eq:lambda_update_c} is satisfied during iteration.

	\subsubsection{Updating RIS Reflection Matrix $\boldsymbol{\Phi}$}
	In this subsection, we present a low-complexity iterative algorithm for updating the RIS reflection matrix $\boldsymbol{\Phi}$. The equivalent channel vector from the BS to user~$k$ can be simplified as $\mathbf{h}^H_k(\mathbf{x})=\boldsymbol{\phi}^H\mathbf{U}_k+\mathbf{u}_k^H$, where $\boldsymbol{\phi}=\operatorname{diag}(\boldsymbol{\Phi}^H)$,  $\mathbf{U}_k=\operatorname{diag}(\mathbf{h}_{r,k}^H)\mathbf{B}^H \boldsymbol{\Sigma} \mathbf{A}$, $\mathbf{u}_k=\mathbf{A}_k^H \boldsymbol{\Sigma}_k \mathbf{1}$. By substituting the simplified channel vector into problem $(\mathbf{P3})$ and fixing other variables, we obtain:
	\begin{subequations}\label{eq:RIS_update}
		\begin{align}
			(\mathbf{P3.2})\: \max_{\boldsymbol{\phi}} \quad 
			& -\boldsymbol{\phi}^{H}\mathbf{M}_{1}\boldsymbol{\phi}
			+ \Re\!\left\{\boldsymbol{\phi}^{H}\left(\mathbf{M}_{2} - 2\mathbf{M}_{3}\right)\right\}
			\\
			\text{s.t.} \quad 
			& \eqref{eq:opt_problem_constraint4},\nonumber\\
			&-\boldsymbol{\phi}^H \mathbf{C}_{1,k} \boldsymbol{\phi}+\Re\{\boldsymbol{\phi}^H(\mathbf{C}_{2,k}-2\mathbf{C}_{3,k})\}\nonumber \\& \quad\quad\quad\quad\quad\quad\quad\quad\quad\quad\quad\geq c_k,\:\forall k,\label{P32rate}\\
			&\sum_{k=1}^{K} r_{c,k}
			\le
			-\boldsymbol{\phi}^{H}\mathbf{N}_{1,k}\boldsymbol{\phi}  \nonumber \\
			&\; + \Re\!\left\{\boldsymbol{\phi}^{H}\left(\mathbf{N}_{2,k} - \mathbf{N}_{3,k}\right)\right\}
			+ d, \: \forall k,\label{eq:RIS_update_b}
		\end{align}
	\end{subequations}	
% 当公式间距太大，需要将公式变得紧凑，可以将所有文字以及公式一起放进$$中，用\\换行，用{=}{+}{-}减少符号两端的间距，用\text标记正文，\text中的文本如果太长需要换行，需要将换行的内容单独放入一个\text。用~表示空格。
$\text{where}~ \mathbf{C}_{1,k}{=}\mathbf{U}_{k}(|\epsilon_{k}|^{2}\sum_{i=1}^{K}\mathbf{w}_{i}\mathbf{w}_{i}^{H})\mathbf{U}_{k}^{H}, \mathbf{C}_{2,k}{=}  2\sqrt{1+\mu_{k}}\bar{\epsilon}_{k}\mathbf{U}_{k}\\\mathbf{w}_{k}, \mathbf{C}_{3,k}{=}\mathbf{U}_{k}(|\epsilon_{k}|^{2}\sum_{i=1}^{K}\mathbf{w}_{i}\mathbf{w}_{i}^{H})\mathbf{u}_{k}, \mathbf{N}_{1,k}  {=}\!|v_{k}|^{2}\mathbf{U}_{k}(\mathbf{w}_{c}\mathbf{w}_{c}^{H}{+}\\ \sum_{i=1}^{K}\mathbf{w}_{i}\mathbf{w}_{i}^{H})\mathbf{U}_{k}^{H}, \mathbf{N}_{2,k}{=}2\sqrt{1+\mu_{k}} \bar{v}_{k}\mathbf{U}_{k}\mathbf{w}_{c}, \mathbf{N}_{3,k}  {=}2|v_{k}|^{2}\mathbf{U}_{k}\\(\mathbf{w}_{c}\mathbf{w}_{c}^{H}{-}\sum_{i=1}^{K}\mathbf{w}_{i}\mathbf{w}_{i}^{H})\mathbf{u}_{k}, \mathbf{M}_1{=}\sum_{k=1}^K \mathbf{C}_{1,k},\mathbf{M}_2{=}\sum_{k=1}^K \mathbf{C}_{2,k},\\ \mathbf{M}_3{=}\sum_{k=1}^K \mathbf{C}_{3,k},~d{=}2\sqrt{1+\gamma_k}\Re\{\bar{v}_k \mathbf{u}_k^H \mathbf{w}_c\}{-}|v_k|^2 \mathbf{u}_k^H \sum_i\mathbf{w}_i \\\mathbf{w}_i^H \mathbf{u}_k,~c_k{=}\iota{+}\mu_k{-}\log_2(1{+}\mu_k){-}2\sqrt{1+\mu_k}\Re\{\bar{\epsilon}_k \mathbf{u}_k^H \mathbf{w}_k\}{+}|\epsilon_k|^2\\\mathbf{u}_k^H \sum_i\mathbf{w}_i \mathbf{u}_k.~\text{The dual problem of}~ (\mathbf{P3.2})~\text{is given by:}$
	\begin{subequations}
		\begin{align}
			(\mathbf{P3.2.1})~\max_{\boldsymbol{\xi}}\;\min_{\boldsymbol{\phi}}\;
			 \mathcal{L}(\boldsymbol{\phi},\boldsymbol{\xi},\boldsymbol{\varpi})\quad \text{s.t.}~ \eqref{eq:opt_problem_constraint4} \nonumber
		\end{align}
	\end{subequations}
where $\mathcal{L}(\boldsymbol{\phi},\boldsymbol{\xi},\boldsymbol{\varpi})$ is the Lagrangian function corresponding to ($\mathbf{P3.2}$). 
%=\boldsymbol{\phi}^H\mathbf{M}_1 \boldsymbol{\phi}-\Re\{\boldsymbol{\phi}^H(\mathbf{M}_2-2\mathbf{M_3}) \}+\sum_{i=1}^K \xi_i (\sum_{k=1}^K r_{c,k}+\boldsymbol{\phi}^H \mathbf{N}_{1,i} \boldsymbol{\phi}-\Re\{\boldsymbol{\phi}^H(\mathbf{N}_{2,i}-\mathbf{N}_{3,i}) \}-d)+\sum_{i=1}^K \varpi_i(c_i+\boldsymbol{\phi}^H \mathbf{C}_{1,i} \boldsymbol{\phi}-\Re\{\boldsymbol{\phi}^H(\mathbf{C}_{2,i}-2\mathbf{C}_{3,i})\})$. 
$\boldsymbol{\xi}=[\xi_1, \xi_2,...,\xi_K]^T$ and $\boldsymbol{\varpi}=[\varpi_1,\cdots,\varpi_K]^T$ denote the Lagrange multiplier corresponding to \eqref{eq:RIS_update_b} and \eqref{P32rate} respectively. We employ the dual ascent method to solve this problem. First, the Lagrange multipliers are fixed to update $\boldsymbol{\phi}$. %After some algebraic manipulations, we obtain $\mathcal{L}(\boldsymbol{\phi},\boldsymbol{\xi})=\boldsymbol{\phi}^H(\mathbf{M_1}+\sum_{i=1}^K \xi_i  \mathbf{N}_{1,i})\boldsymbol{\phi}-\Re\{\boldsymbol{\phi}^H(\mathbf{M}_2-2\mathbf{M}_3+\sum_{i=1}^K \xi_i(\mathbf{N}_{2,i}-\mathbf{N}_{3,i})) \}+\sum_{i=1}^K \xi_i(\sum_{k=1}^K r_{c,k}-d)$. 
	Combining with \eqref{eq:opt_problem_constraint4} and Lemma~1 in \cite{8741198}, %$\boldsymbol{\phi}^H (\mathbf{M}_1+\sum_{i=1}^K \xi_i \mathbf{N}_{1,i}+\sum_{i=1}^K \varpi_i\mathbf{C}_{1,i}) \boldsymbol{\phi} \leq \boldsymbol{\phi}^H \lambda_{max}\mathbf{I} \boldsymbol{\phi}+2\Re\{\boldsymbol{\phi}^H(\mathbf{M}_1+\sum_{i=1}^K \xi_i\mathbf{N}_{1,i}+\sum_{i=1}^K \varpi_i \mathbf{C}_{1,i}-\lambda_{max}\mathbf{I})\boldsymbol{\phi}^{[t]} \}$,
	we can derive an upper-bound function in place of the Lagrangian function as $\Re\{\boldsymbol{\phi}^H \mathbf{h} \}$, where $\mathbf{h}=2(\mathbf{M_1}+\sum_{i=1}^K \xi_i \mathbf{N}_{1,i}+\sum_{i=1}^K \varpi_i\mathbf{C}_{1,i}-\lambda_{max} \mathbf{I})\boldsymbol{\phi}^{[t]}-(\mathbf{M}_2-2\mathbf{M}_3+\sum_{i=1}^K \xi_i(\mathbf{N}_{2,i}-\mathbf{N}_{3,i})+\sum_{i=1}^K \varpi_i(\mathbf{C}_{2,i}-2\mathbf{C}_{3,i}))$,  $\lambda_{max}$ is the maximum eigenvalue of matrix $\mathbf{M}_1+\sum_{i=1}^K \xi_i \mathbf{N}_{1,i}+\sum_{i=1}^K \varpi_i \mathbf{C}_{1,i}$. %Therefore, the update formula for \(\boldsymbol{\phi}\) is given by $\boldsymbol{\phi}^{[t+1]}=-e^{j \angle \mathbf{h}}$. 
	In addition, the gradient ascent method is employed to update $\boldsymbol{\xi}$. %Specifically, the gradient $\nabla_{\boldsymbol{\xi}}\, \mathcal{L}(\boldsymbol{\phi}^{[t]}, \boldsymbol{\xi})$ is selected as the update direction of $\boldsymbol{\xi}$. In summary, 
	The update formulas for $\boldsymbol{\xi}$ and $\boldsymbol{\phi}$ are given as follows:
	\begin{figure*}[!t]
		\centering
		\begin{minipage}{0.95\textwidth}
			\begin{equation}\label{psik}
				\Psi_k(\cdot) = \log_2(1+\mu_k) - \mu_k
				+ \underbrace{2\sqrt{1+\mu_k}\,\Re\{\bar{\epsilon}_k\mathbf{h}_k^{H}(\mathbf{x})\mathbf{w}_k\}}_{\Psi_{k,1}}
				- \underbrace{|\epsilon_k|^2\!\left(\sum_{i=1}^{K}\lvert \mathbf{h}_k^{H}(\mathbf{x})\mathbf{w}_i\rvert^2 + \sigma_k^2\right)}_{\Psi_{k,2}}
			\end{equation}
			\begin{equation}\label{tk}
				T_k(\cdot) = \log_2(1+\gamma_k)-\gamma_k
				+2\sqrt{1+\gamma_k}\,\Re\{\bar{v}_k\mathbf{h}_k^{H}(\mathbf{x})\mathbf{w}_c\}
				-|v_k|^2\!\left(\lvert \mathbf{h}_k^{H}(\mathbf{x})\mathbf{w}_c\rvert^2
				+\sum_{i=1}^{K}\lvert \mathbf{h}_k^{H}(\mathbf{x})\mathbf{w}_i\rvert^2+\sigma_k^2\right)
			\end{equation}
			\vspace{-0.7ex}
			\hrule
		\end{minipage}
	\end{figure*}
	\begin{subequations}
		\begin{align}
			\xi_i^{[t+1]}
			&=
			\Biggl[
			\xi_i^{[t]}
			+ \Biggl(
			\sum_{k=1}^{K} r_{c,k}
			+ (\boldsymbol{\phi}^{[t]})^H\mathbf{N}_{1,i}\boldsymbol{\phi}^{[t]} \nonumber  \\
			&- \Re\!\Bigl\{
			(\boldsymbol{\phi}^{[t]})^{H}(\mathbf{N}_{2,i}-\mathbf{N}_{3,i})
			\Bigr\}
			- d
			\Biggr)\tau
			\Biggr]^{+},
			\label{eq:update_a}\\
			\varpi_i^{[t+1]}&=\Biggl[\varpi_i^{[t]}+\Biggl(c_i+(\boldsymbol{\phi}^{[t]})^H \mathbf{C}_{1,i}\boldsymbol{\phi}^{[t]} \nonumber \\&-\Re\Bigl\{(\boldsymbol{\phi}^{[t]})^H(\mathbf{C}_{2,i}-2\mathbf{C}_{3,i})\Bigr\} \Biggr)\tau\Biggr]^+,\\
			\boldsymbol{\Phi}^{[t+1]}
			&= \operatorname{diag}(-e^{j \angle \mathbf{h}}).
			\label{eq:update_b}
		\end{align}
	\end{subequations}
	where $\tau$ is the step size for gradient ascent and can be set to a small fixed value. $[\cdot]^+ = \max\{0, \cdot\}$ is the projection function for the Lagrange multiplier $\boldsymbol{\xi}$.

	\begin{figure*}[!t]  % [!t] 强制置于页面顶部；如果是双栏 IEEE 论文，figure* 可横跨两栏
		\centering
		
		\begin{subfigure}[b]{0.31\textwidth}
			\centering
			\includegraphics[width=\textwidth, keepaspectratio, trim=0 0 0 0, clip]{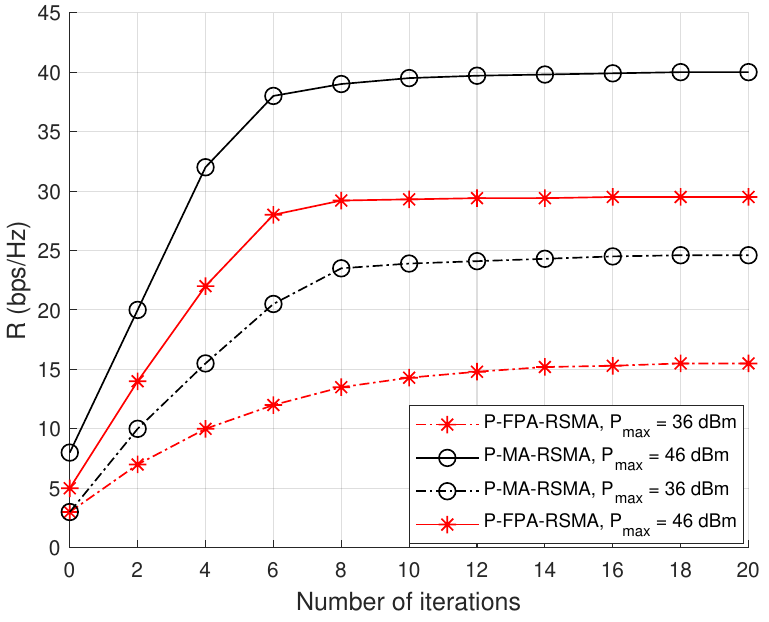}
			\caption{Iteration number versus sum rate}
			\label{fig:convergence}
		\end{subfigure}
		\hfill
		\begin{subfigure}[b]{0.31\textwidth}  % 宽度可微调为 0.30~0.33
			\centering
			\includegraphics[width=\textwidth, keepaspectratio, trim=0 0 0 0, clip]{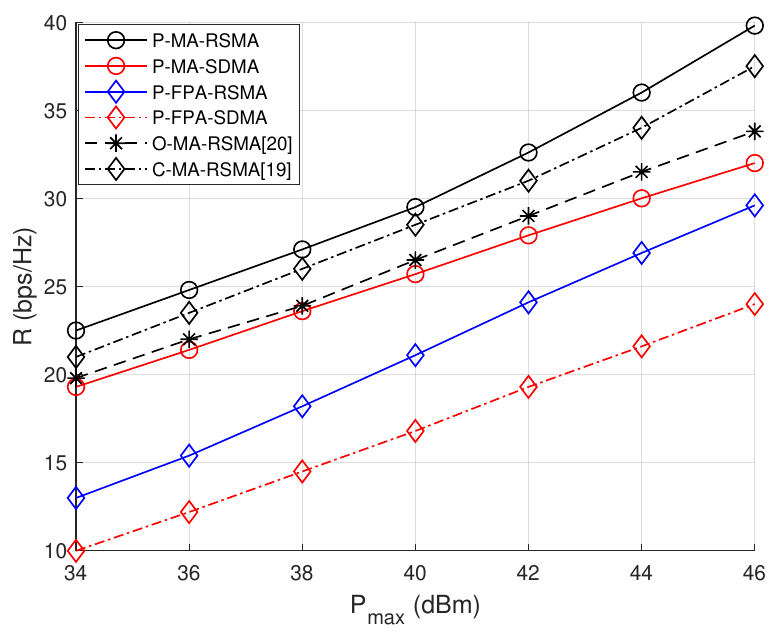}
			\caption{Power budget versus sum rate}
			\label{fig:Performance}
		\end{subfigure}
		\hfill  % 三个子图之间均匀留空
		\begin{subfigure}[b]{0.31\textwidth}
			\centering
			\includegraphics[width=\textwidth, keepaspectratio, trim=0 0 0 0, clip]{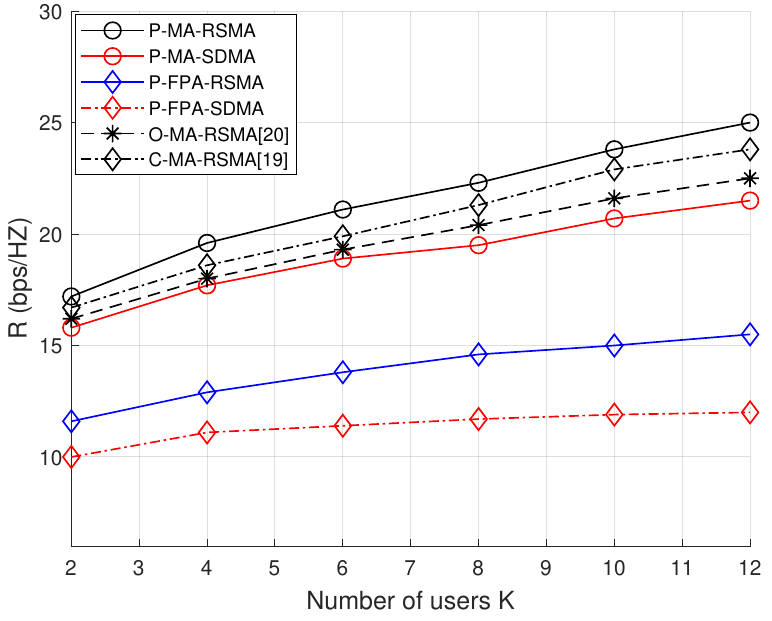}
			\caption{Number of users versus sum rate}
			\label{fig:region}
		\end{subfigure}
		\caption{Performance comparison under different conditions}  
		\label{fig:overall}
		\end{figure*}

	\subsubsection{Updating MA Positions}
	With other variables fixed, we can get the subproblem as follows:
	\begin{subequations}
		\begin{align}
			(\mathbf{P3.3}) \; \max_{\mathbf{x}} \ & \Psi(\mathbf{x})=\sum_{k=1}^K \Psi_k(\mathbf{x}) \\
			\text{s.t. } \ & T_k(\mathbf{x}) \ge y, \ \forall k \\
			& \eqref{region},\ \eqref{eq:opt_problem_constraint5}, \eqref{P3rate}\nonumber
		\end{align}
	\end{subequations}

	Considering the non-convex nature of this problem, we employ the gradient ascent method to update the positions of the MA. The gradient of the objective function with respect to $x_m$ is denoted as $\nabla_{x_m} \Psi(\mathbf{x})$. Accordingly, the position of the $m$-th MA is updated as follows:
	\begin{equation}
		x_m^{[t+1]} = x_m^{[t]} + \alpha (\nabla_{x_m} \Psi_1(\mathbf{x})+\nabla_{x_m} \Psi_2(\mathbf{x})), \forall m
	\end{equation}
	where $\Psi_1(\mathbf{x})=\sum_{k=1}^K \Psi_{k,1}(\mathbf{x})$ and $\Psi_2(\mathbf{x})=\sum_{k=1}^K \Psi_{k,2}(\mathbf{x})$. $\alpha$ denotes the step size of the gradient ascent iteration. Even if a feasible initial solution is selected, the gradient ascent procedure may still violate the feasible region during the iteration process. 
	In such cases, the step size $\alpha$ is reduced to $0.5\alpha$ to ensure that the updated position remains within the feasible set $R_f$, where $R_f$ is the feasible region for problem $(\mathbf{P3.3})$. The gradients of $\Psi_1(\mathbf{x})$ and $\Psi_2(\mathbf{x})$ %can be calculated by (31), where $\mathbf{a}^H=\mathbf{h}_{r,k}^H \Phi \mathbf{B}^H, \mathbf{b}^H=\mathbf{1}^H \Sigma^k$, and $(\cdot)_{m}$ represents the $m^{th}$ element of vector $(\cdot)$. The gradient $\nabla_{\mathbf{x}_m} \Psi_2(\mathbf{x})$ exhibits a similar structure and 
	are omitted for brevity. Also, gradient ascent algorithms for updating MA positions have been introduced in many articles like the \textbf{Algorithm~2} in \cite{11161312}.Therefore, we do not elaborate on them here. The overall algorithm for solving problem $(\mathbf{P1})$ is displayed in \textbf{Algorithm~1}.
	
	% \begin{figure*}[!t]  % [!t] 强制置于页面顶部；如果是双栏 IEEE 论文，figure* 可横跨两栏
	% 	\centering
		
	% 	\begin{subfigure}[b]{0.31\textwidth}
	% 		\centering
	% 		\includegraphics[width=\textwidth, keepaspectratio, trim=0 0 0 0, clip]{result3.pdf}
	% 		\caption{Iteration number versus sum rate}
	% 		\label{fig:convergence}
	% 	\end{subfigure}
	% 	\hfill
	% 	\begin{subfigure}[b]{0.31\textwidth}  % 宽度可微调为 0.30~0.33
	% 		\centering
	% 		\includegraphics[width=\textwidth, keepaspectratio, trim=0 0 0 0, clip]{result1.pdf}
	% 		\caption{Power budget versus sum rate}
	% 		\label{fig:Performance}
	% 	\end{subfigure}
	% 	\hfill  % 三个子图之间均匀留空
	% 	\begin{subfigure}[b]{0.31\textwidth}
	% 		\centering
	% 		\includegraphics[width=\textwidth, keepaspectratio, trim=0 0 0 0, clip]{result2.pdf}
	% 		\caption{Number of users versus sum rate}
	% 		\label{fig:region}
	% 	\end{subfigure}
	% 	\caption{Performance comparison under different conditions}  
	% 	\label{fig:overall}
	% 	\end{figure*}	

	\subsection{Complexity Analysis}
	% The computational complexity of gradient ascent for updating MA positions is approximately $\mathcal{O}(N_G M^2 K^2 L_t^2)$, 
	% where $N_G$ denotes the total number of iterations in the gradient ascent procedure. 
	% In \textbf{Algorithm~1}, the complexities of updating the FP parameters and the Lagrange multipliers 
	% are $\mathcal{O}(8M)$ and $\mathcal{O}(2M(K+1)+K)$, respectively. 
	% During the beamforming matrix update, LU decomposition is be employed to reduce 
	% the inversion complexity in which case only one LU factorization is required, 
	% and the subsequent inverse computations incur a complexity of $\mathcal{O}(M^2)$. 
	% In summary, assuming that the total number of outer iterations is $N_T$, 
	% the overall computational complexity of the proposed algorithm can be approximately expressed as $\mathcal{O}\big(M^3 + N_T(8M + 2M(K+1) + K + M^2 + N_G M^2 K^2 L_t^2)\big)$.
	
	The computational complexity of gradient ascent for updating MA positions is approximately $\mathcal{O}(N_G M^2 K^2 L_t^2)$, where $N_G$ is the total number of gradient ascent iterations. In \textbf{Algorithm~1}, the complexities of updating the FP parameters and Lagrange multipliers are $\mathcal{O}(8M)$ and $\mathcal{O}(2M(K+1)+K)$, respectively. For the beamforming matrix update, LU decomposition is employed to reduce inversion complexity, requiring only one LU factorization, while the subsequent inverse computations incur a complexity of $\mathcal{O}(M^2)$. Therefore, assuming the total number of outer iterations is $N_T$, the overall computational complexity of the proposed algorithm is approximately $\mathcal{O}\big(M^3 + N_T(8M + 2M(K+1) + K + M^2 + N_G M^2 K^2 L_t^2)\big)$.

	% \begin{algorithm}[t]
	% 	\caption{Gradient Descent for Solving Problem (P2.3)}
	% 	\label{alg:gd-p23}
	% 	\begin{algorithmic}[1]
	% 		\STATE \textbf{Initialization:} Initialize $\mathbf{x}^{(l)} \in \mathcal{R}_f$.
	% 		Set iteration index $l=0$, objective value $\Psi^{(l)}=0$, and threshold $\epsilon>0$.
	% 		\STATE Compute $\Psi^{(l+1)}$.
	% 		\REPEAT
	% 		\STATE Update $\mathbf{x}^{(l+1)}$ by (32).
	% 		\IF{$\mathbf{x}^{(l+1)} \notin \mathcal{R}_f$}
	% 		\STATE $\alpha = 0.5 \alpha $.
	% 		\ELSE
	% 		\STATE $\Psi^{(l)} = \Psi^{(l+1)}$.
	% 		\STATE $l = l + 1$.
	% 		\STATE Compute $\Psi^{(l+1)}$.
	% 		\ENDIF
	% 		\UNTIL{$|\Psi^{(l+1)} - \Psi^{(l)}| \le \epsilon$ or $l \ge l_{\max}$}
	% 		\STATE \textbf{Output:} $\mathbf{x}^{(l+1)}$.
	% 	\end{algorithmic}
	% \end{algorithm}
	
	\begin{algorithm}[t]
		\caption{AO for Solving Problem $(\mathbf{P1})$}
		\label{alg:AO-P1}
		\begin{algorithmic}[1]
			\STATE \textbf{Initialization:} Set 
			$\{\boldsymbol{\lambda}^{[t]}, \kappa^{[t]}, \boldsymbol{\eta}^{[t],} \boldsymbol{\mu}^{[t]}, \boldsymbol{\epsilon}^{[t]}, 
			\boldsymbol{v}^{[t]}, \boldsymbol{\gamma}^{[t]}, \mathbf{W}^{[t]},$
			$\mathbf{x}^{[t]}, \boldsymbol{\Phi}^{[t]}, \boldsymbol{\xi}^{[t]},\boldsymbol{\varpi}^{[t]}, \{r_{c,k}^{[t]}\}\}$.
			Set iteration index $t = 0$, maximum iteration number $t_{\max}$, and threshold $\epsilon$. Calculate $\{R^{[t]}, \{R_{c,k}^{[t]}\},\{T_k^{[t]} \}\}$ by \eqref{eq:opt_problem_obj}, \eqref{RCK}, \eqref{tk} respectively.
			
			\REPEAT
			\STATE Update $\boldsymbol{\mu}^{[t+1]}$, $\boldsymbol{\epsilon}^{[t+1]}$, 
			$\boldsymbol{v}^{[t+1]}$, $\boldsymbol{\gamma}^{[t+1]}$ by (15)$\sim$ (18) in \cite{11161312}.
			\STATE Update $\boldsymbol{\lambda}^{[t+1]}$, $\kappa^{[t+1]}$, $\boldsymbol{\eta}^{[t+1]}$ by \eqref{eq:lambda_update_a}$\sim$\eqref{eq:lambda_update_c}.
			\STATE Update $\mathbf{W}^{[t+1]}$ by \eqref{eq:KKT_a} and \eqref{eq:KKT_b}.
			\STATE Calculate $\{R_{c,k}^{[t+1]}\}$ and update 
			$\{r_{c,k}^{[t+1]}\}$ by $\{r_{c,k}^{[t+1]}\} = \frac{1}{K} \min\{R_{c,k}^{[t+1]}\}$, $\forall k$.
			\STATE Update $\boldsymbol{\xi}^{[t+1]}$,$\boldsymbol{\varpi}^{[t+1]}$ and $\boldsymbol{\Phi}^{[t+1]}$ by \eqref{eq:update_a} and \eqref{eq:update_b}.
			\STATE Update $\mathbf{x}^{[t+1]}$ by \textbf{Algorithm 2} in \cite{11161312}.
			\STATE Calculate $R^{[t+1]}$, $\{T_k^{[t+1]}\}$ and set $t = t + 1$.
			\UNTIL{$|R^{[t+1]} - R^{[t]}| < \epsilon$ or $t \ge t_{\max}$}
			\STATE \textbf{Output:}  $R^{[t+1]}$.
		\end{algorithmic}
	\end{algorithm}

\section{Numerical Results}
In this section, we evaluate the performance of the proposed MA-assisted RSMA-RIS algorithm (P-MA-RSMA) and compare it with the CFGS-based MA-RSMA scheme (C-MA-RSMA)~\cite{11161312} and the ORIS-RSMA scheme (O-MA-RSMA)~\cite{qayyum2025power}. We also include the MA-aided SDMA-RIS scheme (P-MA-SDMA) and the FPA-based RSMA-RIS (P-FPA-RSMA) and SDMA-RIS (P-FPA-SDMA) schemes for comparison.

Unless otherwise specified, the simulation parameters are set as follows. The numbers of transmit and receive paths are $L_t=L_r=L=4$, and the elevation and azimuth angles of all paths are uniformly distributed over $[0,\pi]$. The BS employs $M=16$ antennas, and the RIS has $N=32$ reflecting elements. The carrier frequency is $f=3\times10^9$ Hz~\cite{11161312}, and the wavelength is $\lambda=\frac{c}{f}$, where $c$ is the speed of light. The user rate threshold is set to $\iota=1$ bit/s. Each movable antenna can move within $[-6\lambda,6\lambda]$, with a minimum adjacent spacing of $\lambda/2$. The noise power is $-80$ dBm. The BS array center and RIS are located at $(0\,\text{m},0\,\text{m})$ and $(12\,\text{m},16\,\text{m})$, respectively, while users are uniformly distributed in $[20\,\text{m},40\,\text{m}]\times[0\,\text{m},-20\,\text{m}]$. Due to obstacles, no LoS path exists between the BS and users, whereas the BS--RIS and RIS--user links are LoS. Accordingly, the diagonal entries of the BS--user fading matrix $\boldsymbol{\Sigma}_k$ are modeled as $\boldsymbol{\Sigma}_k[i,i]\sim\mathcal{CN}\!\left(0,\frac{1}{p+1}C_0\left(\frac{d_0}{d_k}\right)^{\alpha_1}/L\right)$, $\forall k,\forall i$~\cite{wu2025movable}$,$ where $C_0=\left(\frac{\lambda}{4\pi}\right)^2$, $d_0$ is the unit reference distance, and $d_k$ is the distance between the BS and user $k$. For the BS--RIS channel $\boldsymbol{\Sigma}$, the diagonal entries are given by $\boldsymbol{\Sigma}[1,1]\sim\mathcal{CN}\!\left(0,\frac{p}{p+1}C_0\left(\frac{d_0}{d_{br}}\right)^{\alpha_c}\right)$ and $\boldsymbol{\Sigma}[i,i]\sim\mathcal{CN}\!\left(0,\frac{1}{p+1}C_0\left(\frac{d_0}{d_{br}}\right)^{\alpha_c}/(L-1)\right)$, $\forall i\neq1$, where the Rician factor is $p=0.5$, $c\in\{2,3\}$ corresponds to the BS--RIS and RIS--user links, respectively, and $d_{br}$ is the BS--RIS distance. The path-loss exponents are set to $\alpha_1=3.5$, $\alpha_2=2.5$, and $\alpha_3=2.5$~\cite{wu2025movable}. %In Fig.~\ref{fig:overall}, MA-RSMA, MA-SDMA, FPA-RSMA, and FPA-SDMA denote the MA-aided RSMA-RIS, MA-aided SDMA-RIS, FPA-based RSMA-RIS, and FPA-based SDMA-RIS schemes, respectively.

First, Fig.~\ref{fig:overall}(a) analyzes the evolution of the RSMA sum rate versus the iteration index under two transmit-power constraints. Across all four schemes, the curves converge in approximately 10 iterations and then remain stable, with the higher-power setting reaching a larger steady-state sum rate.

Then, we investigate how the sum rate \(R\) scales with the maximum transmit power \(P_{\max}\) in Fig.~\ref{fig:overall}(b). The number of users is set to $K = 12$. It can be observed that the performance of the proposed algorithm in this work is superior to the algorithms in \cite{11161312,qayyum2025power}. As expected, \(R\) increases monotonically with \(P_{\max}\), and the slope for RSMA is larger than that for SDMA. Specifically, as the power budget increases from 34 dBm to 46 dBm, the sum rate of MA-RSMA increases by approximately 81.8\%, while that of MA-SDMA increases by only approximately 64.1\%. This is because RSMA affords higher degrees of freedom (DoF) than SDMA\cite{mao2022rate}. Furthermore, it can be observed that both SDMA-RIS and RSMA-RIS systems achieve significant performance gains with the incorporation of MA. Specifically, MA provides approximately a 35.6\% performance enhancement for the RSMA system, compared to about 33.3\% for the SDMA system at $46$dBm.%, indicating that MAs shape the channel in a manner more favorable to RSMA.

In Fig.~\ref{fig:overall}(c), we examine the relationship between the number of users \(K\) and the sum rate. The power budget is set to $36$dBm.
Overall, the sum rate increases with \(K\). %, but the increments diminish as \(K\) becomes larger.
%Moreover, the MA-assisted RSMA--RIS curve exhibits a steeper slope than the alternatives, indicating that the MA-induced improvement is more pronounced for RSMA than for SDMA as \(K\) grows. 
As the number of users increases from 2 to 12, the sum rate of MA-RSMA increases by approximately 47.1\%, while that of FPA-RSMA increases by only approximately 29.2\%.
 %Moreover, as observed from the figure, this performance gap continues to widen with the increasing number of users. Intuitively, when \(K\) increases, MA must reshape the user channels toward near-orthogonality to substantially benefit SDMA, whereas RSMA does not require such stringent orthogonality to achieve strong performance.

	\section{Conclusion} 
	In this work, we investigate the MA assisted RSMA–RIS downlink MISO system and maximize the sum rate via joint optimization of the transmit beamforming matrix, the common-rate allocation, the RIS reflection matrix, and the MA positions. We develop an efficient solution based on a FP reformulation to solve this problem. Numerical results demonstrate that, our proposed algorithm can effectively solve the initial probelm and prove that MA can improve the performance of RSMA–RIS system.

\bibliographystyle{IEEEtran}
\bibliography{paper1}	
% \bibitem{b1} G. Eason, B. Noble, and I. N. Sneddon, ``On certain integrals of Lipschitz-Hankel type involving products of Bessel functions,'' Phil. Trans. Roy. Soc. London, vol. A247, pp. 529--551, April 1955.
% \bibitem{b2} J. Clerk Maxwell, A Treatise on Electricity and Magnetism, 3rd ed., vol. 2. Oxford: Clarendon, 1892, pp.68--73.
% \bibitem{b3} I. S. Jacobs and C. P. Bean, ``Fine particles, thin films and exchange anisotropy,'' in Magnetism, vol. III, G. T. Rado and H. Suhl, Eds. New York: Academic, 1963, pp. 271--350.
% \bibitem{b4} K. Elissa, ``Title of paper if known,'' unpublished.
% \bibitem{b5} R. Nicole, ``Title of paper with only first word capitalized,'' J. Name Stand. Abbrev., in press.
% \bibitem{b6} Y. Yorozu, M. Hirano, K. Oka, and Y. Tagawa, ``Electron spectroscopy studies on magneto-optical media and plastic substrate interface,'' IEEE Transl. J. Magn. Japan, vol. 2, pp. 740--741, August 1987 [Digests 9th Annual Conf. Magnetics Japan, p. 301, 1982].
% \bibitem{b7} M. Young, The Technical Writer's Handbook. Mill Valley, CA: University Science, 1989.
% \end{thebibliography}
% \vspace{12pt}
% \color{red}
% IEEE conference templates contain guidance text for composing and formatting conference papers. Please ensure that all template text is removed from your conference paper prior to submission to the conference. Failure to remove the template text from your paper may result in your paper not being published.

\end{document}